\documentclass[12pt]{article}
\usepackage{amsmath}
\usepackage{amsfonts}
\usepackage{mathptm}
\usepackage{epsfig}
\usepackage{pst-plot,epsf}
\newcommand{\beq}{\begin{equation}}
\newcommand{\eeq}{\end{equation}}
\newcommand{\bea}{\begin{eqnarray}}
\newcommand{\eea}{\end{eqnarray}}
\newcommand{\beas}{\begin{eqnarray*}}
\newcommand{\eeas}{\end{eqnarray*}}

\newcommand{\epm}{e^+e^-}
\newcommand{\gv}{\mbox{GeV}}

\begin{document}
\thispagestyle{empty}
\voffset -2cm
\begin{flushright}
%TP-USl-01/00\\
March 2023\\
Revised:\\
May/July 2023\\
\vspace*{1.cm}
\end{flushright}
\begin{center}
{\Large\bf PSGen, a generator of phase space parameterizations for 
the multichannel Monte Carlo integration}\\
\vspace*{1.5cm}
%----------------------------
Karol Ko\l odziej\footnote{E-mail: karol.kolodziej@us.edu.pl}\\[1cm]
{\small\it
Institute of Physics, University of Silesia\\ 
ul. 75 Pu\l ku Piechoty 1, PL-41500 Chorz\'ow, Poland}\\
\vspace*{1.5cm}
{\bf Abstract}\\
\end{center}
PSGen is a new general purpose Fortran program which has been written to facilitate 
the Monte Carlo phase space integration of the 
$S$ matrix element of any $2 \to n$ scattering process, 
with $n=2,...,9$, provided by the user. 
The program is written in Fortran 90/95. It uses a new very fast algorithm that 
automatically generates calls to Fortran subroutines containing 
different phase space parameterizations of the considered class of processes.
The parameterizations take into account mappings of poles due 
to the Feynman propagators of unstable heavy particles decaying into 2 or 3 
on shell final state particles according to predefined patterns, possible 
single or double $t$-channel poles and peaks due to one on shell photon or gluon 
radiation. The individual subroutines are organized in a single multichannel
kinematics subroutine which can be easily called while computing the phase 
space 
integral of the $S$ matrix element as a function of generated particle four momenta, 
in either the leading or higher orders of the perturbation 
series. The particle four momenta can be used in a quadruple precision version, 
if necessary. 

\vfill

\newpage

\section{Introduction}
Projects of the High-Luminosity Large Hadron Collider (HL-LHC)~\cite{HL-LHC} and
electron--positron colliders as
the Future Circular Collider (FCC--ee)~\cite{FCC} and Compact Linear 
Collider (CLIC)~\cite{CLIC} at CERN, the International Linear Collider 
(ILC)~\cite{ILC} in Japan, or the Circular Electron--Positron Collider 
(CEPC)~\cite{CEPC} in China, offer a wealth of new possibilities
to test various aspects of the theory of fundamental interactions.
Questions about the non-Abelian nature of gauge symmetry group and
the mechanism of the symmetry breaking can be directly addressed in processes 
of a few heavy particles 
production at a time, such as the top-quark pair production, possibly associated
with the Higgs or a heavy electroweak gauge boson, or processes of a few
heavy bosons production at a time. 
In order to explore the nature of the heavy particles
interaction, the corresponding multiparticle decay products of them
must be studied in detail, including their distributions and spin correlations. 
Reactions with multiparticle final states must also be taken into account if 
one wants 
to determine precisely hadronic contributions to the vacuum polarization through
dispersion relations from measurements of the ratio 
$R=\sigma(\epm\to \text{hadrons})/\sigma(\epm\to \mu^+\mu^-)$ at the centre of mass
energies below  the $J/\psi$ production threshold.
The hadronic contributions to the vacuum polarization are the major factor which
influences precision of theoretical predictions for the muon $g-2$ anomaly 
%which is being mesured at Fermilab 
and plays an important role in the evolution of 
the fine structure constant $\alpha(Q^2)$ from the Thomson limit to high energy 
scales. 

In order to fully exploit physical information contained in reactions with the 
multiparticle final states it is necessary to perform numerical integration 
over a multidimensional phase space of the corresponding squared modulus of 
matrix elements, often involving several dozen thousand or even several
hundred thousand amplitudes of the Feynman diagrams. The amplitudes include peaks, mostly due to denominators of the Feynman 
propagators, which must be mapped out in order to obtain reliable results of the
integration. This goal can be in practice obtained only 
within the  multichannel Monte Carlo (MC) approach, with
the corresponding integration routine being generated in a fully automatic way. 

Most of the processes of interest in the high energy accelerators can be handled with
existing general purpose programs, such as:
{\tt MadGraph/MadEvent/HELAS} \cite{MADGRAPH}, 
{\tt CompHEP/CalcHEP} \cite{CompHEP}, 
{\tt ALPGEN} \cite{ALPGEN},
{\tt HELAC-PHEGAS} \cite{HELAC-PHEGAS}, 
{\tt SHERPA/Comix} \cite{SHERPA/Comix},
{\tt O'Mega/Whizard} \cite{OMEGA/WHIZARD}, or 
{\tt carlomat}~\cite{carlomat}, \cite{carlomat2}, \cite{carlomat3}, 
\cite{carlomat4}. 
Some programs, as 
{\tt FeynArts/FormCalc} \cite{FeynForm}, 
{\tt GRACE} \cite{GRACE}, {\tt MadGraph5\_aMC@NLO} \cite{MADGRAPH-NLO}, 
{\tt SHERPA~2.2} \cite{sherpa-NLO} and
{\tt HELAC-NLO} \cite{HELAC-NLO},
enable automatic calculation of the NLO EW or QCD corrections. Most of
those programs also offer a possibility of integrating the generated matrix elements
over the phase space, even if it is multidimensional. 

However, some interesting issues, as {\it e.g.} the above mentioned determination 
of hadronic 
contributions to the vacuum polarization, require dedicated studies of 
multiparticle reactions within some effective models.
In such cases, researchers usually spend a lot of time to program necessary 
matrix elements by themselves and then they must invest yet more time to 
prepare a routine for reliable phase space integration.
The present work tries to meet their needs providing a new
tool, called PSGen, which automatically generates a stand-alone Fortran 90/95 
subroutine which, if called with random arguments by any MC integration routine, 
delivers the corresponding particle four momenta calculated for one selected 
phase space paremeterization together with properly normalized differential 
volume element of the multidimensional phase space.

The paper is organized as follows. Basics of program PSGen are 
described in Section~2, sample results are presented in Section~3 and 
preparation for running and program usage are described in Section~4.

\section{Basics of PSGen}
PSGen is a Fortran 90/95 program which automatically generates calls to 
kinematics subroutines of the user defined process. 
The generated subroutines and auxiliary files are moved to the target directory, 
{\it i.e.} user defined destination directory,
where they are organized in a subroutine containing 
the differential multichannel phase space volume parameterization that can be easily 
called by any MC integration routine.
\subsection{Generation of kinematics routines}
The core part of PSGen is subroutine {\tt genps(nfspt)}. It contains an algorithm 
for generating calls to handwritten subroutines containing different 
phase space parameterizations, further referred to as kinematics routines, 
for a given
number of the final state particles {\tt nfspt}. The latter is determined 
automatically from the 
character variable {\tt process}, which is defined by the user in {\tt PSGen.f} 
and transferred to subroutine {\tt read\_process(process)}. The algorithm 
is based on user defined
patterns which are collected in a data file {\tt genps.dat}. Each pattern consists of
one line that contains 
the following data: a number of the final state particles, their names and, 
after a colon, the mass and width of the intermediate particle(s) they are 
coupled to. For example, in the current 
version of file {\tt genps.dat}, among others, there are the following lines:\\
{\tt 2 u u\~{}  : mg,zero},\\
{\tt 2 e- e+   : mz,gamz}.\\
The first line means that a pair of the final state quarks $u\bar{u}$ couples to the 
intermediate gluon of mass {\tt mg} and width {\tt zero} and the second line says 
that the $e^-e^+$ pair couples
to the $Z$ boson of mass {\tt mz} and width {\tt gamz}. There are also entries in
{\tt genps.dat} which look like the following one:\\
{\tt 3 b\~{} d u\~{}  : mw,gamw,mt,gamt}.\\
It consists of 3 final state particles $\bar{b}d\bar{u}$ which couple to  two 
intermediate particles: the $d\bar{u}$-quark pair couples 
the $W$ boson of mass {\tt mw} and width {\tt gamw} and the $W$ boson and 
$\bar{b}$-quark couple to the top quark of mass {\tt mt} and width {\tt gamt}. 
If the number of particles is 0, then the whole line is treated as a comment, 
{\it e.g.} the line\\
{\tt 0 quark-quark-gluon:}\\
is a comment.
The names of particles, their masses and widths in file {\tt genps.dat} must conform 
with those listed in file {\tt particles.dat}, where in addition to the name and width 
also some other characteristics 
of each particle are given, namely two integers equal 1 or 0 each, and the type 
of the particle in the form of {\tt character(1)} variable at the end of each data 
line. The first integer
specifies whether the particle couples $(=1)$ or not $(=0)$ to the photon, 
the second specifies if it couples to the gluon and the one character 
variable specifies the type of particle, 
{\it i.e.}, {\tt n} stands for a neutrino, 
{\tt l} for a charged lepton, {\tt q} for quark, etc. 

After the process has been defined and a few flags, which will be explained
below, have been specified, the program {\tt PSGen} makes a call to the subroutine 
{\tt read\_process(process)}. It reads the initial and final state particles from
character variable {\tt process} and
checks if all the particles are contained in data file {\tt particles.dat}
and whether they couple to the photon or gluon.
The patterns listed in file {\tt genps.dat} are used to generate 
the subroutine {\tt kinschnl}, which comprises calls to kinematics routines 
containing mappings smoothing peaks due to the Feynman propagators of the 
intermediate particles. The subroutine {\tt read\_process(process)}
also checks if there are $t$-channel poles in the process, or if the final state 
contains a photon or a gluon. If it is so, then {\tt genps(nfspt)}
will also generate the file {\tt tchcalls.f}, which comprises calls 
to kinematics routines containing mappings of the $t$-channel poles, 
or the subroutine {\tt kingchnl}, which contain calls to subroutines with mappings 
of poles due to radiation of the external photon or gluon. File {\tt tchcalls.f}
is included in handwritten subroutine {\tt kintchnl} which is located 
in the target directory. 

Calls to subroutines {\tt kinschnl}, {\tt kintchnl} and {\tt kingchnl} are all 
included in the automatically generated subroutine {\tt kincls}, unless
the user decides otherwise by choosing appropriate values of flags {\tt itchnl} 
or {\tt igchnl} in program {\tt PSGen}. If {\tt itchnl} ({\tt igchnl}) is set to 0, 
then a 
call to {\tt kintchnl} ({\tt kingchnl}) in {\tt kincls} is cancelled. Obviously, the
calls to them are not made, if there are no $t$-channel poles or the external
photon or gluon are not present in the process. Yet one more flag {\tt iquadp}  
is present in program {\tt PSGen}. If {\tt iquadp=1} then the quadruple precision 
for denominators of the Feynman propagators, the particle masses and four momenta 
is used, otherwise the double precision arithmetic is used.
\subsection{The kinematics routine}
In the current distribution of the program, all the generated routines and auxiliary 
files, which are also written in Fortran 90/95,
are shifted to the directory {\tt ../mc\_computation},
where they are used by the kinematics routine
\bea
\label{psgen}
\text{\tt subroutine psgen(ikin,ecm,x1,x2,x,ndim,flux,dlips)}.
\eea
Subroutine (\ref{psgen}) utilizes the multichannel MC approach, 
{\it i.e.}, it combines calls to different phase space parameterizations containing
the mappings discussed in Section~2.1
in a single phase space parameterization. It is self-consistent
in a sense that it can be easily called by any program which integrates the $S$ 
matrix element in either 
the leading or higher orders of the perturbation series. Its automatically 
generated ingredients can be used
in a quadruple precision version, if necessary. 

In the current distribution, subroutine (\ref{psgen}) is called from a template 
{\tt function cs\_sect(x,ndim)}, 
that is integrated by a template {\tt program PSGen\_test\_mpi} with the use of
MC integration routine {\tt carlos}, see {\it e.g.} \cite{carlomat4}. 
Both the routine from which (\ref{psgen})
is called and the main MC integration program must include the command\\
{\tt use kinparams},\\
where {\tt module kinparams} is automatically created at the stage of code generation. 
Moreover, the main integration program must include the command\\
{\tt call param\_trans(unit)},\\
which should be located below the command that opens the output file associated 
with the same {\tt unit} number and before the first call to the actual MC integration
routine used.

The dummy arguments of (\ref{psgen}) are the following:
\begin{itemize}
\item {\tt ikin = 1,2,...,nkin}, where {\tt nkin} is the number of the generated
calls to the kinematics channels; parameter {\tt nkin} is defined in 
the automatically generated {\tt module kinparams},
\item {\tt ecm} is the user defined centre of mass energy,
\item {\tt x1,x2} are the beam energy fractions carried by the initial state 
particles, {\it i.e.},
if {\tt x1=x2=1} then the initial state particles scatter at fixed energy {\tt ecm},
\item {\tt x} is an array of random numbers of dimension {\tt ndim}, 
with {\tt x(ndim)}
being delivered by the MC integration routine actually used,
\item {\tt flux} and {\tt dlips} are, respectively, the initial flux factor and the
differential phase space element, both calculated in (\ref{psgen}).
\end{itemize}
The particle four momenta computed in (\ref{psgen}) for a given value of {\tt ikin} 
are returned in the {\tt module fourmom} which is also created
at the stage of code generation. The module is used in (\ref{psgen}) and must also 
be used wherever the user wants to refer to the particle four momenta.
\subsection{Handwritten kinematics subroutines}
A number of kinematics subroutines corresponding to $2,3,...,9$ final 
state particles have been written and tested. Each of them calculates a volume 
of the Lorentz invariant 
phase space volume element as generally defined in Eq. (2) of \cite{carlomat4}
and the set of the final state particle four momenta corresponding to the random
arguments {\tt x(ndim)} they are called with. 
If a double $t$-channel pole is present in the process then it is mapped out 
in a way described in Section~3.1 of \cite{carlomat4}. Several new subroutines 
have been written which map the single $t$-channel poles in a similar way and
subroutines which map out peaks due to the photon or gluon radiation with 
transformations described in Section~3.1 of \cite{carlomat4}. Lists of dummy arguments
of all those subroutines take into account patterns defined in the file 
{\tt genps.dat},
as discussed in Section~2.1. It may happen, however, that for some new user defined
patterns in {\tt genps.dat} or for some processes for which the program has 
not been tested yet, new kinematics subroutines will be necessary. Note that, 
as most of those handwritten subroutines contain the Fortran kind type 
parameters which are set at 
the stage of code generation, they  must be recompiled each time the MC code
is generated anew. However, this is not a problem at all, as the compilation
usually takes few seconds.

The user can easily add by hand a call to an own made kinematics subroutine 
by modifying the automatically generated subroutine {\tt kincls} which collects 
calls to all kinematics subroutines of different type. 
The own made kinematics subroutine should be written in the similar way as
the automatically generated subroutines {\tt kinschnl} or {\tt kingchnl}.
The instruction where the
call should be put is contained in {\tt kincls.f}. Note, that the number of 
kinematics
channels given by parameter {\tt nkin} in {\tt kinparams.f} must then be adjusted 
appropriately by hand according to the following formula
\beas
{\tt nkin}={\tt nkinag}+{\tt nkinua},
\eeas
where {\tt nkinag} is the number of the kinematics channels automatically 
generated and {\tt nkinua} is the number of channels added by the user.

All the physical input parameters are defined in the {\tt module~inprms\_ps}, 
located in the directory {\tt mc\_computation}, where in 
particular numerical values of all the particle masses and widths introduced in files
{\tt genps.dat} and {\tt particles.dat} must be specified.

\section{Sample results}
To illustrate the accuracy of the new phase space generator, we compare results
of the MC integration of the leading order (LO) cross sections of a few reactions 
which have been obtained with the phase parameterization based on topologies 
of the Feynman
diagrams, as automatically generated by {\tt carlomat\_4.4} \cite{carlomat4}, 
the most recent update of {\tt carlomat\_4.0}, and 
the phase space parameterization obtained with {\tt PSGen} and used for 
integration of the LO matrix elements generated by {\tt carlomat\_4.4}. The phase 
space integration range has been restricted by the following cuts on the charged 
lepton--beam, photon--beam, and photon--lepton angles and energies of the charged 
lepton and photon: 
\bea
\label{cuts}
5^{\circ} < \theta(l,\mathrm{beam}), \theta (\gamma,\mathrm{beam}) < 175^{\circ},\quad 
\theta( \gamma,l) > 5^{\circ}, \quad E_{l}> 5\;\gv, \quad E_{\gamma}> 1\;\gv.
\eea
The cross sections are shown in Table~\ref{tab:css}. 
\begin{table}[ht]
\begin{center}
\begin{tabular}{lcccc}
$\qquad$ Reaction & \multicolumn{2}{c}{$\sigma(0.5\;{\rm TeV})$} 
&\multicolumn{2}{c}{$\sigma(1\;{\rm TeV})$} \\[1.5mm]
\hline
\hline
 & {\tt carlomat\_4.0} & {\tt PSGen} &{\tt carlomat\_4.0} & {\tt PSGen}\\[1.5mm]
$\epm\to b \mu^+\nu_{\mu}\bar{b}\mu^-\bar{\nu}_{\mu}$ & 6.565(4)~fb & 6.593(3)~fb 
           & 2.332(8)  & 2.375(2)~fb \\[1.5mm] % carlo: 8'.41'', PSGen: 6'03''
                                % 10 times 2000000
$\epm\to b e^+\nu_{e}\bar{b}\mu^-\bar{\nu}_{\mu}$ & 6.624(4)~fb & 6.622(4)~fb 
           & 2.621(11)  & 2.594(6)~fb \\[1.5mm] % carlo: 7'52'', PSGen: 5'21''
                                % 10 times 2000000
$\epm\to b \mu^+\nu_{\mu}\bar{b}\mu^-\bar{\nu}_{\mu}\gamma$ & 1.602(5)~fb & 1.551(11)~fb 
           & 0.722(4)  & 0.715(5)~fb \\[1.5mm] % carlo: 43'.18'', PSGen: 31'08''
                                % 10 times 2000000
\end{tabular}
\caption{LO cross sections at $\sqrt{s}=0.5$~TeV and $\sqrt{s}=1$~TeV 
with cuts of (\ref{cuts}). 
Uncertainties of the last digits are given in parentheses.\label{tab:css}}
\end{center}
\end{table}
The accuracy, in terms of one standard deviation of the
MC integration, which is given for every entry in the parentheses, is comparable for
both integrations. However, {\tt PSGen} generates much less kinematics
channels than {\tt carlomat\_4.4}. This results in much shorter time of the code
generation and compilation and shorter program execution time.

\section{Preparation for running and program usage}
%
%Other minor changes to the program, including corrections of a few 
%bugs are described in a {\tt readme} file. 

Program {\tt PSGen} is distributed as a single {\tt tar.gz} archive
{\tt PSGen.tgz} which can be downloaded from:
http://kk.us.edu.pl/PSGen.html.
When extracted with the command\\
{\tt tar -xzvf PSGen.tgz}\\
it will create directory {\tt PSGen\_1.0} with sub directories:
{\tt code\_generation}, {\tt mc\_computation} and 
{\tt test\_output}.

The preparation for running requires the following steps
\begin{itemize}
\item Choose a Fortran 90/95 compiler in a {\tt makefile}
of {\tt code\_generation} and possibly change the target directory to which 
the generated files should be moved from {\tt ../mc\_computation/} to a directory
of your choice. Note that the target directory must include subroutine (\ref{psgen})
and all other handwritten files listed in a makefile of {\tt mc\_computation}.
\item Specify the process and required
options in {\tt PSGen.f} and execute the command\\ 
{\tt make code}\\ 
from the command line in {\tt code\_generation}.
\end{itemize}
As already mentioned in Section~2.2, the current distribution of {\tt PSGen} contains 
a template program {\tt PSGen\_test\_mpi} which allows to perform the MC integration, 
utilizing the Message Passing Interface (MPI), of a template function 
{\tt cs\_sect(x,ndim)}, both located in directory 
{\tt mc\_computation}. Function {\tt cs\_sect(x,ndim)} calls the 
kinematics routine (\ref{psgen}). As the conversion constant and the matrix 
element are both set to 1 in {\tt cs\_sect(x,ndim)}, the integral is the phase space 
volume of the considered process, restricted by kinematics cuts, 
at the centre 
of mass energies defined in {\tt PSGen\_test\_mpi} by array {\tt aecm(ne)}.
The cuts can be defined in the subroutine {\tt define\_cuts} and imposed by 
setting {\tt icuts=1} in {\tt PSGen\_test\_mpi}.
\begin{itemize}
\item Edit {\tt makefile} of {\tt mc\_computation} by defining the number 
of processor cores on which the MC program will be executed, currently 
{\tt n\_cores:=4}, and choose the MPI Fortran
compiler. 
If you choose the Intel Fortran compiler, {\tt FF = mpiifort}, then 
uncomment the line\\
{\tt       include 'mpif.h'}\\
and comment the line\\
{\tt       use mpi}
in {\tt PSGen\_test\_mpi.f} and do vice versa if you choose the GNU Fortran compiler
{\tt FF = mpif90} or {\tt FF = mpifort}.
\end{itemize}

The template program can be run with the command\\
{\tt make mc}\\
executed in the directory {\tt mc\_computation}.
The MC integration is performed using the multichannel approach with integration
weights adjusted anew after every iteration, as described in 
Section~2 of \cite{carlomat4}.
Obviously, the templates can be used to calculate the cross section of the user 
defined process, if a call to user's own subroutine calculating the corresponding 
matrix element
at the four momenta calculated by (\ref{psgen}) is made 
and the conversion constant is appropriately taken into account function in
{\tt cs\_sect(x,ndim)}.

The basic output of the MC run is written in file {\tt tot\_0\_name}, where 
{\tt name} is created automatically if the assignment for character variable\\
{\tt nicknm='auto'}\\
in {\tt PSGen.f} is not changed to arbitrary user's defined name. The output
of other MPI processes is written to files {\tt tot\_i\_name}, with {\tt i}
being the MPI process identification number. Current directory 
{\tt test\_output} contains results for the phase space volume of reactions
listed in Table~\ref{tab:css} at $\sqrt{s}=500$~GeV and $\sqrt{s}=1$~TeV
with cuts (\ref{cuts}), 
calculated in 10 iterations of $2\times 10^5$ calls each. Files {\tt tot\_0\_...}
contain output of the MPI root process, {\it i.e.} the combined output of the four
MPI processes corresponding to {\tt n\_cores:=4}. The actual values
of physical parameters and cuts used in the computation are listed in the beginning 
of each output file. Note, that the cross sections contained in them differ 
from those of Table~\ref{tab:css}, as they were calculated with the squared module 
of the corresponding matrix element set to 1, {\em i.e.} {\tt mat2=1}.
The parameters can be redefined in {\tt inprms\_ps.f} and cuts as well as 
default settings for the MC integration
can be changed in {\tt PSGen\_test\_mpi.f}. There are also files
{\tt memos\_i\_...} present in the directory. They may contain some warnings 
concerning potential numerical inaccuracies which may occur during 
the run of the corresponding MPI process, with {\tt i=0,1,2,3}. Some of them
may be empty, if no such warning messages have been generated. 

Whenever the Fortran compiler is changed, or a
compiled program is transferred to another computer with a different
processor, all the object and module files should be deleted 
by executing the commands:\\
{\tt rm *.o}\\
{\tt rm *.mod}\\
and the command {\tt make mc} should be executed anew. It is also recommended 
to do so every time the physical parameters or cuts have been changed, just in case
the routine dependencies in {\tt makefile} are not all properly taken into account.

\end{document}